\documentclass[prb]{revtex4}
\usepackage{amsmath}
\usepackage{amsfonts}

\begin{document}

\title{Invariance of the Noether charge}
\author{ Z.~K.~Silagadze}
\email{silagadze@inp.nsk.su}
\affiliation{Budker Institute of Nuclear Physics SB RAS and Novosibirsk State
University, 630 090, Novosibirsk, Russia }

\date{\today}

\begin{abstract}
Surprisingly, an interesting property of the Noether charge that it is 
by itself invariant under the corresponding symmetry transformation is 
never discussed in quantum field theory or classical mechanics textbooks
we have checked. This property is also almost never mentioned in  
articles devoted to Noether's theorem. Nevertheless, to prove
this property in the context of Lagrangian formalism is not quite trivial
and the proof, outlined in this article, can constitute an useful and 
interesting exercise for students.
\end{abstract}

\maketitle

\section{Introduction}
Noether's theorem \cite{1,1A,1B} is a fundamental result which establishes 
a connection between continuous symmetries and conservation laws. Both
concepts play a central role in modern physics. It is not surprising, 
therefore, that it is discussed in many quantum and classical field theory 
textbooks \cite{2,3,4,5,6,7,8,9,10,11,12,13,14}, as well as in some classical
mechanics textbooks of various levels of sophistication \cite{15,16,17,18,19,
20,21,22,23,24,25,26}. It is surprising, however, that it is hard to find an 
answer in the quoted literature to the natural question of how these conserved 
Noether charges are affected by the corresponding symmetry transformations.
Moreover, neither Hill's well-known review \cite{27} nor various pedagogical 
expositions of the Noether's theorem \cite{28,29,30,31,32,33,34} discuss this 
question.

Our intuitive understanding is that symmetry is a property of the system
to remain unchanged under some kind of transformation. Noether charges are 
among important characteristics of the system which determine its physical
state. Therefore a natural expectation is that Noether charges should not 
be changed under the corresponding symmetry transformations. This is indeed 
the case. However the  invariance property of the Noether charge is  
``rather hard to prove'' in Lagrangian formalism \cite{35}. In the context of
classical mechanics, the proofs were given  by Lutzky, for the case of 
a system with one degree of freedom \cite{36}, and by  Sarlet and Cantrijn 
for the general case \cite{35} (see also \cite{37,37A}).

In the field theory context, the invariance of the Noether charge follows
from a more general mathematical result first proved by Khamitova \cite{38}
(after it was conjectured by Nail Ibragimov). Later Khamitova's result 
describing the action of symmetries on conservation laws was reformulated 
in somewhat different language as Proposition 5.64 in  Olver's book \cite{39}.

The aim of this note is to give a pedagogical exposition of this interesting 
property of the Noether charge in the frameworks of both classical mechanics 
and field theory.

\section{Noether theorem in classical mechanics}
Let us consider a classical mechanical system whose dynamics is determined 
by Hamilton's variational principle  
\begin{equation}
\delta\int\limits_{t_1}^{t_2}L(t,q,\dot q)\,dt=0
\label{eq1}
\end{equation}
yielding the Euler-Lagrange equations
\begin{equation}
\frac{d}{dt}\left(\frac{\partial L}{\partial \dot{q}^i}\right)=
\frac{\partial L}{\partial q^i}.
\label{eq2}
\end{equation}
Here $q$ and $\dot q$ are shorthand notations for generalized coordinates
$q=(q^1(t),q^2(t),\ldots,q^n(t))$ and the corresponding velocities
$\dot q=(\dot q^1(t),\dot q^2(t),\ldots,\dot q^n(t))$. An infinitesimal
transformation
\begin{equation}
t^\prime=t+\epsilon\,\tau(t,q),\;\;\;q^{\prime\,i}(t^\prime)=q^{i}(t)+
\epsilon\,\xi^i(t,q)
\label{eq3}
\end{equation}
is said to be a symmetry of the system considered if it leaves invariant 
the Euler-Lagrange equations of motion. A sufficient condition that the
transformation (\ref{eq3}) is a symmetry is provided by the existence of 
such function $K(t,q)$ that up to the first order in the transformation 
parameter $\epsilon\ll 1$ the following identity holds true:
\begin{equation}
L\left(t^\prime(t),q^\prime(t^\prime(t)),\frac{dq^\prime(t^\prime)}{dt^\prime}
(t)\right)\frac{dt^\prime(t)}{dt}=L(t,q(t),\dot q(t))+
\epsilon\,\frac{dK(t,q)}{dt},
\label{eq4}
\end{equation}
where (we use Einstein summation convention that repeated indexes are 
implicitly summed over)
\begin{equation}
\frac{dK(t,q)}{dt}=\frac{\partial K(t,q)}{\partial t}+\dot{q}^i\,
\frac{\partial K(t,q)}{\partial q^i}.
\label{eq5}
\end{equation}
Indeed, in this case the new action integral 
$$S^\prime=\int\limits_{t^\prime_1}^{t^\prime_2}L\left(t^\prime,q^\prime
(t^\prime),\frac{dq^\prime(t^\prime)}{dt^\prime}\right) dt^\prime$$
remains quasi-invariant:
\begin{equation}
S^\prime=\int\limits_{t_1}^{t_2}L\left(t^\prime(t),q^\prime
(t^\prime(t)),\frac{dq^\prime(t^\prime)}{dt^\prime}(t)\right)
\frac{dt^\prime(t)}{dt}\,dt=S+\epsilon\,[K(t_2,q(t_2))-K(t_1,q(t_1))],
\label{eq6}
\end{equation}
and we will have $\delta S^\prime=\delta S+\epsilon\,\delta [K(t_2,q(t_2))-
K(t_1,q(t_1))]=0$, if $\delta S=0$, because it is assumed in the Hamilton's 
variational principle that variations of the generalized coordinates vanish 
at the initial and final points (at $t=t_1$ and $t=t_2$ respectively).

The velocity transformation law under (\ref{eq3}) is the following
\begin{equation}
\frac{dq^{\prime\,i}(t^\prime)}{dt^\prime}=\frac{dq^i+\epsilon\,d\xi^i}
{dt+\epsilon\,d\tau}\approx \dot{q}^i+\epsilon\,(\dot{\xi}^i-\dot{q}^i\,
\dot{\tau}),
\label{eq7}
\end{equation}
where a dot denotes total derivative with respect to time $t$. For example,
$$\dot \tau=\frac{\partial \tau}{\partial t}+\dot{q}^i\,
\frac{\partial \tau}{\partial q^i}.$$
Therefore, we can introduce the generator of the transformation (\ref{eq3}), 
\begin{equation}
\hat G=\tau(t,q)\,\frac{\partial}{\partial t}+\xi^i(t,q)\,
\frac{\partial}{\partial q^i}+\eta^i(t,q,\dot q)\,\frac{\partial}
{\partial \dot q^i},
\label{eq8}
\end{equation}
with
\begin{equation}
\eta^i(t,q,\dot q)=\dot{\xi}^i-\dot{q}^i\,\dot{\tau},
\label{eq9}
\end{equation}
so that for any function $f(t,q,\dot q)$ its variation under the  
transformation (\ref{eq3}) is
\begin{equation} 
\delta f=f(t^\prime,q^\prime(t^\prime),dq^\prime(t^\prime)/dt^\prime)-
f(t,q(t),\dot q(t))=\epsilon\,\hat G(f).
\label{eq10}
\end{equation}
Sometimes it is necessary to extend (\ref{eq8}) by including higher 
derivatives. For example, in light of (\ref{eq7}) we have
\begin{equation}
\frac{d^2q^{\prime\,i}(t^\prime)}{dt^{\prime\,2}}=\frac{d(dq^{\prime\,i}
(t^\prime)/dt^\prime)}{dt^\prime}\approx \frac{d\dot q^i+\epsilon\, d\eta^i}
{dt+\epsilon\, d\tau}\approx \ddot q^i(t)+\epsilon\,\zeta^i(t,q,\dot q,
\ddot q),
\label{eq11}
\end{equation}
where
\begin{equation}
\zeta^i(t,q,\dot q,\ddot q)=\dot{\eta}^i-\ddot{q}^i\,\dot{\tau}.
\label{eq12}
\end{equation}
Therefore the prolongation of the operator (\ref{eq8}) on the space
$(t,q,\dot q,\ddot q)$ has the form (the same symbol will be used both for 
the transformation operator and any of its prolongations)
\begin{equation}
\hat G=\tau(t,q)\,\frac{\partial}{\partial t}+\xi^i(t,q)\,
\frac{\partial}{\partial q^i}+\eta^i(t,q,\dot q)\,\frac{\partial}
{\partial \dot q^i}+\zeta^i(t,q,\dot q,\ddot q)\,\frac{\partial}
{\partial \ddot q^i}.
\label{eq13}
\end{equation} 
Introducing the Lie characteristic function
\begin{equation}
\sigma^i(t,q,\dot q)=\xi^i(t,q)-\dot q^i\, \tau(t,q),
\label{eq14}
\end{equation}
and using
\begin{equation}
\eta^i-\tau\,\ddot q^i=\dot \sigma^i,\;\;\; \zeta^i-\tau\,\dddot q^i=
\ddot \sigma^i,
\label{eq15}
\end{equation}
along with
\begin{equation}
\frac{\partial}{\partial t}=\frac{d}{dt}-\dot q^i\,\frac{\partial}
{\partial q^i}-\ddot q^i\,\frac{\partial}{\partial \dot q^i}-\dddot q^i\,
\frac{\partial}{\partial \ddot q^i},
\label{eq16}
\end{equation}
the generator (\ref{eq13}) can be rewritten in the form
\begin{equation}
\hat G=\tau\,\frac{d}{d t}+\sigma^i\,\frac{\partial}{\partial q^i}+
\dot \sigma^i\,\frac{\partial}{\partial \dot q^i}+\ddot \sigma^i
\,\frac{\partial}{\partial \ddot q^i}=\tau\,\frac{d}{d t}+\hat L_B.
\label{eq17}
\end{equation} 
Here we have introduced the canonical Lie-B\"{a}cklund operator \cite{40}
\begin{equation}
\hat L_B=\sigma^i\,\frac{\partial}{\partial q^i}+
\dot \sigma^i\,\frac{\partial}{\partial \dot q^i}+\ddot \sigma^i
\,\frac{\partial}{\partial \ddot q^i}.
\label{eq18}
\end{equation} 
Although we shall not particularly need this fact here, the same simple 
pattern continues to hold for prolongations to higher jet spaces 
(by including higher derivatives of $q^i$) \cite{40} and sometimes it is 
technically more convenient to work with completely prolonged operators. 
For example, let us show that the total time derivative operator commutes 
with the canonical Lie-B\"{a}cklund operator \cite{40}. Assuming that $k$ 
and $l$ indexes run from zero to infinity, we write
\begin{equation}
\left[\frac{d}{dt},\,\hat L_B\right]=\left[\frac{d}{dt},\,\sigma^{i\,(l)}\,
\frac{\partial}{\partial q^{i\,(l)}}\right]=\sigma^{i\,(l+1)}\,\frac{
\partial}{\partial q^{i\,(l)}}+\sigma^{i\,(l)}\left[\frac{d}{dt},\,\frac{
\partial}{\partial q^{i\,(l)}}\right].
\label{eq19}
\end{equation} 
On the other hand,
\begin{equation}
\left[\frac{d}{dt},\,\frac{\partial}{\partial q^{i\,(l)}}\right]=
\left[\frac{\partial}{\partial t}+q^{j\,(k+1)}\,\frac{\partial}{\partial 
q^{j\,(k)}},\,\frac{\partial}{\partial q^{i\,(l)}}\right]=-\frac
{\partial q^{j\,(k+1)}}{\partial q^{i\,(l)}}\,\frac{\partial}{\partial 
q^{j\,(k)}}=-\delta_l^{k+1}\,\frac{\partial}{\partial q^{i\,(k)}},
\label{eq20}
\end{equation} 
where $\delta_l^{k+1}$ denotes the Kronecker delta function. Substituting
this into (\ref{eq19}), we get
\begin{equation}
\left[\frac{d}{dt},\,\hat L_B\right]=\sigma^{i\,(l+1)}\,\frac{
\partial}{\partial q^{i\,(l)}}-\sigma^{i\,(l)}\,\delta_l^{k+1}\,\frac{\partial}
{\partial q^{i\,(k)}}=\sigma^{i\,(l+1)}\frac{\partial}{\partial q^{i\,(l)}}-
\sigma^{i\,(k+1)}\,\frac{\partial}{\partial q^{i\,(k)}}=0.
\label{eq21}
\end{equation} 
The canonical Lie-B\"{a}cklund operator determines the so called vertical
variation
\begin{equation} 
\bar \delta f=f(t,q^\prime(t),\dot q^\prime(t))-f(t,q(t),\dot q(t))=
\epsilon\,\hat L_B(f),
\label{eq22}
\end{equation}
which is caused solely by the changes in functional forms of generalized
coordinates and their derivatives. In particular
\begin{equation} 
\bar \delta q^i=q^{\prime\,i}(t)-q^i(t)=q^{\prime\,i}(t^\prime)-
q^i(t)-[q^{\prime\,i}(t^\prime)-q^{\prime\,i}(t)]\approx \epsilon\,(\xi^i-
\dot q^i\tau)=\epsilon\,\sigma^i=\epsilon\,\hat L_B(q^i).
\label{eq23}
\end{equation}
Using
\begin{equation}
\frac{dt^\prime}{dt}=1+\epsilon\,\dot\tau(t,q,\dot q),
\label{eq24}
\end{equation}
and
\begin{equation}
\epsilon\,\dot\tau(t,q,\dot q)L\left(t^\prime(t),q^\prime(t^\prime(t)),
\frac{dq^\prime(t^\prime)}{dt^\prime}(t)\right)\approx
\epsilon\,\dot\tau(t,q,\dot q)L(t,q,\dot q),
\label{eq25}
\end{equation}
we get from (\ref{eq4})
\begin{equation}
\hat G(L)=\tau\dot L+\hat L_B(L)=\dot K-\dot\tau L,
\label{eq26}
\end{equation}
which implies
\begin{equation}
\hat L_B(L)=\frac{d}{dt}\,(K-\tau L).
\label{eq27}
\end{equation}
On the other hand,
\begin{equation}
\hat L_B(L)=\sigma^i\,\frac{\partial L}{\partial q^i}+\dot\sigma^i\,\frac
{\partial L}{\partial \dot q^i}=\sigma^i\left (\frac{\partial L}{\partial q^i}
-\frac{d}{dt}\,\frac{\partial L}{\partial \dot q^i}\right)+\frac{d}{dt}
\left(\sigma^i\,\frac{\partial L}{\partial \dot q^i}\right)=\sigma^i\,\frac{
\delta L}{\delta q^i}+\frac{d}{dt}\left(\sigma^i\,\frac{\partial L}{\partial 
\dot q^i}\right),
\label{eq28}
\end{equation}
where we have introduced the Euler-Lagrange operator (variational derivative)
\begin{equation}
\frac{\delta }{\delta q^i}=\frac{\partial }{\partial q^i}-\frac{d}{dt}\,
\frac{\partial }{\partial \dot q^i}.
\label{eq29}
\end{equation}
Its prolongations to higher jet spaces can be read from the 
expression \cite{37A}
\begin{equation}
\frac{\delta }{\delta q^i}=\frac{\partial }{\partial q^i}+\sum\limits_{l\ge 
1}(-1)^l\,\frac{d^l}{dt^l}\,\frac{\partial }{\partial q^{i\,(l)}}.
\label{eq30}
\end{equation}
Equations (\ref{eq27}) and (\ref{eq28}) imply the validity of the so-called 
Rund-Trautman identity \cite{41,42}
\begin{equation}
\frac{d}{dt}\left (K-\tau L-\sigma^i\,\frac{\partial L}{\partial \dot q^i}
\right)=\sigma^i\,\frac{\delta L}{\delta q^i},
\label{eq31}
\end{equation}
from which the Noether theorem (in fact Noether's first theorem) readily 
follows: for every continues symmetry transformation (\ref{eq3}) there exits
a conserved Noether charge
\begin{equation}
Q=K-\tau L-\sigma^i\,\frac{\partial L}{\partial \dot q^i}.
\label{eq32}
\end{equation}
Indeed, (\ref{eq31}) and the Euler-Lagrange equations (\ref{eq2}) guarantee 
that $\dot Q=0$.

Sometimes $K(t,q)$ is called the Bessel-Hagen function (see, for example, 
\cite{33}), because Noether in her celebrated paper considered only $K=0$ 
case and more general case of symmetries up to divergence were introduced 
later by Erich Bessel-Hagen \cite{43}. However the problem was suggested 
to Bessel-Hagen by Noether herself \cite{43,44}.

\section{Invariance of the Noether charge in classical mechanics}
The Noether charge (\ref{eq32}) can be rewritten in the following way
\begin{equation}
Q=K-\hat N(L),
\label{eq33}
\end{equation} 
where
\begin{equation}
\hat N=\tau+\sigma^i\,\frac{\partial}{\partial \dot q^i}+\left(\dot\sigma^i
-\sigma^i\,\frac{d}{dt}\right )\frac{\partial}{\partial \ddot q^i},
\label{eq34}
\end{equation}
is the Ibragimov operator (in the more general form, it was introduced by
Ibragimov \cite{37A,40} under the name Noether operator. We find it more 
appropriate to call it Ibragimov operator).

The last term in (\ref{eq34}) has no effect when working in the first jet 
space $(t,q,\dot q)$ and that's why (\ref{eq32}) and (\ref{eq33}) are 
equivalent on the $(t,q,\dot q)$ space. So, at first sight, its introduction
is superfluous. However this extra term will prove to be very useful as we
are going now to show. Using (\ref{eq20}), we get
\begin{equation}
\frac{\partial}{\partial \dot q^i}\,\frac{d}{dt}=\left [\frac{\partial}
{\partial \dot q^i},\,\frac{d}{dt}\right ]+\frac{d}{dt}\,\frac{\partial}
{\partial \dot q^i}=\frac{\partial}{\partial q^i}+\frac{d}{dt}\,\frac{\partial}
{\partial \dot q^i},
\label{eq35}
\end{equation}
and analogously
\begin{equation}
\frac{\partial}{\partial \ddot q^i}\,\frac{d}{dt}=\frac{\partial}
{\partial \dot q^i}+\frac{d}{dt}\,\frac{\partial}{\partial \ddot q^i}.
\label{eq36}
\end{equation}
Therefore
\begin{equation}
\hat N\,\frac{d}{dt}=\tau\,\frac{d}{dt}+\sigma^i\left(\frac{\partial}
{\partial q^i}+\frac{d}{dt}\,\frac{\partial}{\partial \dot q^i}\right)+
\left(\dot\sigma^i-\sigma^i\,\frac{d}{dt}\right )\left(\frac{\partial}
{\partial \dot q^i}+\frac{d}{dt}\,\frac{\partial}{\partial \ddot q^i}\right),
\label{eq37}
\end{equation}
which simplifies to
\begin{equation}
\hat N\,\frac{d}{dt}=\hat G+\left(\dot\sigma^i-\sigma^i\,\frac{d}{dt}\right )
\frac{d}{dt}\,\frac{\partial}{\partial \ddot q^i}.
\label{eq38}
\end{equation}
The last term can be neglected in the $(t,q,\dot q)$ space and we get the 
following very useful identity (with above mentioned more general definition 
of $\hat N$ it can be made strictly valid in all jet spaces \cite{37A})
\begin{equation}
\hat G=\hat N\,\frac{d}{dt}.
\label{eq39}
\end{equation}
Let us calculate the commutator
$$\left [\frac{d}{dt},\,\hat N\right]=\left [\frac{d}{dt},\,
\tau+\sigma^i\,\frac{\partial}{\partial \dot q^i}+\left(\dot\sigma^i
-\sigma^i\,\frac{d}{dt}\right )\frac{\partial}{\partial \ddot q^i}\right ].$$
Neglecting the terms which are irrelevant in the first jet 
space $(t,q,\dot q)$, we get
\begin{equation}
\left [\frac{d}{dt},\,\hat N\right]=\dot\tau+\dot\sigma^i\,\frac{\partial}
{\partial \dot q^i}-\sigma^i\,\frac{\partial}{\partial q^i}-\left(\dot\sigma^i
-\sigma^i\,\frac{d}{dt}\right )\,\frac{\partial}{\partial \dot q^i}=
\dot\tau-\sigma^i\,\frac{\delta}{\delta q^i}.
\label{eq40}
\end{equation}
Therefore
\begin{equation}
[\hat G,\,\hat N]=\left [\hat N\,\frac{d}{dt},\,\hat N\right ]=\hat N\,\left [
\frac{d}{dt},\,\hat N\right ]=\hat N\,\left (\dot\tau-\sigma^i\,\frac{\delta}
{\delta q^i}\right).
\label{eq41}
\end{equation}
Now we are well equipped to prove the invariance of the Noether charge. Indeed
we have
\begin{equation}
\hat G(Q)=\hat G(K-\hat N(L))=\hat G(K)-\hat G\,\hat N(L).
\label{eq42}
\end{equation}
But
\begin{equation}
\hat G\,\hat N(L)=[\hat G,\,\hat N](L)+\hat N\,\hat G (L),
\label{eq43}
\end{equation}
which after using (\ref{eq26}), (\ref{eq41}) and the Euler-Lagrange equations
becomes
\begin{equation}
\hat G\,\hat N(L)=\hat N(\dot\tau\,L)+\hat N(\dot K-\dot\tau\,L)=
\hat N(\dot K).
\label{eq44}
\end{equation}
Substituting this result into (\ref{eq42}) and using (\ref{eq39}), we get 
finally
\begin{equation}
\hat G(Q)=\hat G(K)- \hat N(\dot K)=\hat N\,\frac{d}{dt}\,(K)-\hat N(\dot K)=0.
\label{eq45}
\end{equation}
As we see the Noether charge is indeed invariant under the corresponding 
symmetry transformation (\ref{eq3}), as it should be according to our 
intuitive understanding of symmetry.

\section{Noether theorem in classical field theory}
Next we consider $n$-component classical field $u_a(x)$, $a=1,\ldots n$
in the Minkowski space-time with coordinates $x^\mu$. It is assumed that the 
classical dynamics of the field is governed by the action principle
\begin{equation}
\delta S=\delta\int\limits_\Omega dx\,{\cal L}(x,u,u_{,\,\mu})=0.
\label{eq46}
\end{equation}
Here $\Omega=[t_1,\,t_2]\times {\mathbb R}^3$ is the space-time domain and 
comma indicates differentiation with respect to $x$:
\begin{equation}
u_{a,\,\mu}=\frac{du_a(x)}{dx^\mu}.
\label{eq47}
\end{equation}
We shall proceed as much as possible in analogy with the classical mechanical
case. In particular, the transformation
\begin{equation}
x^{\prime\,\mu}=x^\mu+\epsilon\,\tau^\mu(x,u),\;\;\;
u^\prime_a(x^\prime)=u_a(x)+\epsilon\,\xi_a(x,u)
\label{eq48}
\end{equation}
is a symmetry if the following holds true
\begin{equation}
{\cal L}\left(x^\prime(x),u^\prime(x^\prime(x)),\frac{du^\prime
(x^\prime)}{dx^\prime}
(x)\right)J(x)={\cal L}(x,u(x),u_{,\,\mu}(x))+\epsilon\,K^\mu_{,\,\mu}
\label{eq49}
\end{equation}
for some functions $K^\mu(x,u)$. To avoid a confusion, for such functions
comma denotes total differentiation with respect to the indicated component
of $x$:
\begin{equation}
K^\mu_{,\,\nu}=\frac{dK^\mu}{dx^\nu}=\frac{\partial K^\mu}{\partial x^\nu}+
u_{a,\,\nu}\,\frac{\partial K^\mu}{\partial u_a}.
\label{eq50}
\end{equation}
At last, $J=\mathrm{det}[\partial x^{\prime\,\mu}/\partial x^\nu]$ is the 
Jacobian corresponding to the transformation $x\to x^\prime$.

introducing the generator of the transformation (\ref{eq48}), $\hat G$, and
taking into account that
\begin{equation}
J\approx 1+\epsilon\,\tau^\mu_{,\,\mu},
\label{eq51}
\end{equation}
the symmetry condition (\ref{eq49}) can be rewritten in the form
\begin{equation}
\hat G ({\cal L})=K^\mu_{,\,\mu}-\tau^\mu_{,\,\mu}\,{\cal L}.
\label{eq52}
\end{equation}
Under (\ref{eq48}), the field derivatives transform as follows
\begin{equation}
\frac{du^\prime_a(x^\prime)}{dx^{\prime\,\mu}}=\frac{\partial x^\nu}
{\partial x^{\prime\,\mu}}\,\frac{du^\prime_a(x^\prime)}{dx^\nu}\approx
(\delta^\nu_\mu-\epsilon\,\tau^\nu_{,\,\mu})(u_{a,\,\nu}+\epsilon\,
\xi_{a,\,\nu})\approx u_{a,\,\mu}+\epsilon\,(\xi_{a,\,\nu}-
\tau^\nu_{,\,\mu}\,u_{a,\,\nu}).
\label{eq53}
\end{equation}
Therefore the generator $\hat G$ has the form
\begin{equation}
\hat G=\tau^\mu \,\frac{\partial }{\partial x^\mu}+\xi_a\,\frac{\partial }
{\partial u_a}+\eta_{a\mu}\,\frac{\partial }{\partial u_{a,\,\mu}},
\label{eq54}
\end{equation}
where
\begin{equation}
\eta_{a\mu}=\xi_{a,\,\mu}-\tau^\nu_{,\,\mu}\,u_{a,\,\nu}.
\label{eq55}
\end{equation}
In complete analogy with (\ref{eq14}) and (\ref{eq17}), it is easy to rewrite
the generator $\hat G$ in the form
\begin{equation}
\hat G=\tau^\mu\,\frac{d}{dx^\mu}+\sigma_a\,\frac{\partial}{\partial u_a}+
\sigma_{a,\,\mu}\,\frac{\partial}{\partial u_{a,\,\mu}}=\tau^\mu\,\frac{d}
{dx^\mu}+\hat L_B,
\label{eq56}
\end{equation}
with the Lie characteristic function
\begin{equation}
\sigma_a=\xi_a-\tau^\mu\,u_{a,\,\mu}.
\label{eq57}
\end{equation}
Now we have
\begin{equation}
\hat L_B({\cal L})=\sigma_a\left( \frac{\partial {\cal L}}{\partial u_a}-
\frac{d}{dx^\mu}\,\frac{\partial {\cal L}}{\partial u_{a,\,\mu}}\right)+
\frac{d}{dx^\mu}\left(\sigma_a\,\frac{\partial {\cal L}}{\partial u_{a,\,\mu}}
\right)=\sigma_a\,\frac{\delta {\cal L}}{\delta u_a}+\frac{d}{dx^\mu}\left(
\sigma_a\,\frac{\partial {\cal L}}{\partial u_{a,\,\mu}}\right)
\label{eq58}
\end{equation}
and, in combination with (\ref{eq52}) and (\ref{eq56}), (\ref{eq58}) implies
the validity of the field theoretical version of the Rund-Trautman identity
\begin{equation}
\frac{d}{dx^\mu}\left(K^\mu-\tau^\mu\,{\cal L}-\sigma_a\,\frac{\partial 
{\cal L}}{\partial u_{a,\,\mu}}\right )=\sigma_a\,\frac{\delta {\cal L}}
{\delta u_a}.
\label{eq59}
\end{equation}
Then the Euler-Lagrange equations
\begin{equation}
\frac{\delta {\cal L}}{\delta u_a}=\frac{\partial {\cal L}}{\partial u_a}-
\frac{d}{dx^\mu}\,\frac{\partial {\cal L}}{\partial u_{a,\,\mu}}=0
\label{eq60}
\end{equation}
imply the existence of the conserved (divergence-free) current
\begin{equation}
J^\mu=K^\mu-\tau^\mu\,{\cal L}-\sigma_a\,\frac{\partial {\cal L}}
{\partial u_{a,\,\mu}},\;\;\frac{dJ^\mu}{dx^\mu}=0.
\label{eq61}
\end{equation}
The corresponding conserved Noether charge, associated with the symmetry 
transformation (\ref{eq48}), is
\begin{equation}
Q=\int J^0\,d\vec{x}.
\label{eq62}
\end{equation}
So far, so good. However, unfortunately, here the simple analogy with the 
classical mechanical case ends and we need some extra labor to extend the
proof of invariance of the Noether charge to the field theory case also.

\section{Invariance of the Noether charge in classical field theory}
Let us introduce again the Ibragimov operator
\begin{equation}
\hat N^\mu=\tau^\mu+\sigma_a\,\frac{\partial}{\partial u_{a,\,\mu}},
\label{eq63}
\end{equation}
so that
\begin{equation}
J^\mu=K^\mu-\hat N^\mu({\cal L}).
\label{eq64}
\end{equation}
It is shown in the appendix that, in the first jet space  
$(x^\mu,u_a,u_{a,\,\nu})$, the following commutation relation, which
will play an important role in our arguments below, holds true:
\begin{equation}
[\hat G+\tau^\nu_{,\,\nu},\,\hat N^\mu]=\tau^\mu_{,\,\nu}\,\hat N^\nu.
\label{eq65}
\end{equation}
In fact, for suitably defined $\hat G$ and $\hat N^\mu$, (\ref{eq65}) is
valid in all jet spaces \cite{37A}. Now we use this commutation relation
in the following way. We have
\begin{equation}
\hat G \hat N^\mu({\cal L})=[\hat G+\tau^\nu_{,\,\nu},\,\hat N^\mu]({\cal L})
+\hat N^\mu(\hat G+\tau^\nu_{,\,\nu})({\cal L})-\tau^\nu_{,\,\nu}\hat N^\mu
({\cal L}),
\label{eq66}
\end{equation}
which after using (\ref{eq65}) and (\ref{eq52}) becomes
\begin{equation}
\hat G \hat N^\mu({\cal L})=\tau^\mu_{,\,\nu}\,\hat N^\nu({\cal L})+
\hat N^\mu(K^\nu_{,\,\nu})-\tau^\nu_{,\,\nu}\hat N^\mu({\cal L}).
\label{eq67}
\end{equation}
Let us substitute here $\hat N^\mu({\cal L})=K^\mu-J^\mu$ from (\ref{eq64})
and rearrange the terms. As a result we get \cite{37A}
\begin{equation}
\hat G(J^\mu)+\tau^\nu_{,\,\nu}\,J^\mu-\tau^\mu_{,\,\nu}\,J^\nu=
\hat G(K^\mu)+\tau^\nu_{,\,\nu}\,K^\mu-\tau^\mu_{,\,\nu}\,K^\nu-
\hat N^\mu(K^\nu_{,\,\nu}).
\label{eq68}
\end{equation}
Of course, this is far more complicated result than (\ref{eq45}) and it is
not immediately obvious how it can lead to invariance of the corresponding
Noether charge. Nevertheless (\ref{eq68}) indeed imply this invariance, as
we now will show.

First of all it is necessary to understand what the invariance of the Noether 
charge does mean in the context of field theory. Let $\tilde x=(x^0-\epsilon\,
\tau^0(x,u),\,\vec{x})$, so that after the transformation  (\ref{eq48})
$\tilde x^\prime =(x^0,\,\vec{x}^{\,\prime})$. The Noether charge $Q$ doesn't
depend on time. Therefore
\begin{equation}
Q=\int J^0(x,u(x),u_{,\,\mu}(x))\,d\vec{x}=\int J^0(\tilde{x},u(\tilde{x}),
u_{,\,\mu}(\tilde{x}))\,d\vec{x},
\label{eq69}
\end{equation}
and  after the transformation  (\ref{eq48}) it becomes
\begin{equation}
Q^\prime=\int J^0(\tilde{x}^\prime,u^\prime(\tilde{x}^\prime),
u^\prime_{,\,\mu}(\tilde{x}^\prime))\,d\vec{x}^{\,\prime}=
\int J^0(x,u^\prime(x),u^\prime_{,\,\mu}(x))\,d\vec{x}.
\label{eq70}
\end{equation}
The last equality follows from the fact that $\vec{x}^{\,\prime}$ is a dummy 
variable in  (\ref{eq70}). Therefore, the invariance of the Noether charge,
$Q^\prime=Q$, means that
\begin{equation}
\int\left[J^0(x,u^\prime(x),u^\prime_{,\,\mu}(x))-J^0(x,u(x),u_{,\,\mu}(x))
\right]\,d\vec{x}\approx\epsilon\int\hat L_B(J^0)\,d\vec{x}=0
\label{eq71}
\end{equation}
and we come to the following condition
\begin{equation}
\int\hat L_B(J^0)\,d\vec{x}=0.
\label{eq72}
\end{equation}
Now let us return to (\ref{eq68}) and substitute
$$\hat G=\tau^\nu\,\frac{d}{dx^\nu}+\hat L_B.$$
As a result we get
\begin{equation}
\hat L_B(J^\mu)=\left[\tau^\nu(K^\mu-J^\mu)\right]_{,\,\nu}+
\tau^\mu_{,\,\nu}\,J^\nu-\tau^\mu_{,\,\nu}\,K^\nu+\hat L_B(K^\mu)-
\hat N^\mu(K^\nu_{,\,\nu}),
\label{eq73}
\end{equation}
where we have taken into account that, for example
\begin{equation}
\tau^\nu\,J^\mu_{,\,\nu}+\tau^\nu_{,\,\nu}\,J^\mu=(\tau^\nu J^\mu)_{,\,\nu}.
\label{eq74}
\end{equation}
Next we have
\begin{equation}
\hat L_B(K^\mu)-\hat N^\mu(K^\nu_{,\,\nu})=\sigma_a\,\frac{\partial K^\mu}
{\partial u_a}-\tau^\mu\,K^\nu_{,\,\nu}-\sigma_a\,\frac{\partial 
K^\nu_{,\,\nu}}{\partial u_{a,\,\mu}}.
\label{eq75}
\end{equation}
But
\begin{equation}
K^\nu_{,\,\nu}=\frac{\partial K^\nu}{\partial x^\nu}+u_{b,\,\nu}
\frac{\partial K^\nu}{\partial u_b},
\label{eq76}
\end{equation}
and 
\begin{equation}
\frac{\partial K^\nu_{,\,\nu}}{\partial u_{a,\,\mu}}=\delta^a_b\,\delta^\mu_
\nu\,\frac{\partial K^\nu}{\partial u_b}=\frac{\partial K^\mu}{\partial u_a}.
\label{eq77}
\end{equation}
Therefore
\begin{equation}
\hat L_B(K^\mu)-\hat N^\mu(K^\nu_{,\,\nu})=-\tau^\mu\,K^\nu_{,\,\nu},
\label{eq78}
\end{equation}
and (\ref{eq73}) takes the form
\begin{equation}
\hat L_B(J^\mu)=\left[\tau^\nu(K^\mu-J^\mu)\right]_{,\,\nu}+
\tau^\mu_{,\,\nu}\,J^\nu-(\tau^\mu K^\nu)_{,\,\nu}.
\label{eq79}
\end{equation}
But $J^\nu_{,\,\nu}=0$, as $J^\nu$ is a conserved current. Therefore
\begin{equation}
\tau^\mu_{,\,\nu}\,J^\nu=\tau^\mu_{,\,\nu}\,J^\nu+\tau^\mu\,J^\nu_{,\,\nu}=
(\tau^\mu J^\nu)_{,\,\nu},
\label{eq80}
\end{equation}
and substituting this into (\ref{eq79}) leads to a little miracle:
\begin{equation}
\hat L_B(J^\mu)=\frac{d}{dx^\nu}\left[\tau^\nu(K^\mu-J^\mu)-\tau^\mu(K^\nu-
J^\nu)\right]=\frac{dG^{\mu\nu}}{dx^\nu}.
\label{eq81}
\end{equation}
The fact that 
\begin{equation}
G^{\mu\nu}=\tau^\nu(K^\mu-J^\mu)-\tau^\mu(K^\nu-J^\nu)=
\tau^\mu(J^\nu-K^\nu)-\tau^\nu(J^\mu-K^\mu)
\label{eq82}
\end{equation}
is an antisymmetric tensor plays the crucial role, because then
\begin{equation}
\hat L_B(J^0)=\frac{dG^{0i}}{dx^i},\;\;\;i=1,2,3,
\label{eq83}
\end{equation}
is the total three-dimensional divergence and the validity of (\ref{eq72})
then follows from the Gauss theorem, provided our system is closed, so that 
fields fall sufficiently rapidly at spatial infinity to render the limit of
the resulting surface integral zero.

\section{Concluding remarks}
Noether charge is invariant with respect to the corresponding symmetry 
transformation, as expected. In the context of classical mechanics, the 
initial rather brute-force proof  by Lutzky \cite{36} and by  
Sarlet and Cantrijn \cite{35} can be significantly simplified by using
ideas from \cite{37A}.

In classical field theory, our presentation of this interesting property of
the Noether charge is also based on the results of Ibragimov, Kara and 
Mahomed \cite{37A}, in particular on the commutation relation  (\ref{eq65}).
The crucial relation (\ref{eq81}), from which the invariance of the Noether 
charge follows, is a particular case of a more general result of 
Khamitova \cite{38}. However, the paper \cite{38} is not an easy reading due
to omission of many calculational details and to our knowledge it has not been
used in the context of invariance of the Noether charge in the classical field 
theory.

One more point is worth to be mentioned. In Hamiltonian framework for 
classical mechanics both Noether theorem \cite{17,35,45} and invariance of 
the Noether charge \cite{35} are almost trivial results. Hamiltonian proof of 
the Noether theorem is so simple that it even ``makes one question why the 
statement should be considered an important result'' \cite{17}. Besides, the 
description of symmetry and conserved quantities in Hamiltonian framework is 
more straightforward and powerful than in the Lagrangian framework \cite{45}.
Why should we then bother about  Lagrangian version of the Noether theorem
at all? The answer is simple: because of its important applications in
the field theories of modern physics. ``Since Lagrangian is a relativistic 
invariant in field theory while Hamiltonian is not, the Lagrangian formalism 
has a special advantage over the Hamiltonian formalism in relativistic quantum 
mechanics'' \cite{46}. The quantization of modern gauge theories is also most 
straightforwardly formulated via Lagrangian path integral which is very 
convenient in practical calculations because it gives manifestly Lorentz 
invariant expressions (in covariant gauges) and easily leads to  the Feynman 
rules. 

Hamiltonian point of view is important as it underlines the rich geometrical
ideas behind the Noether's theorem. In such a framework the proof of the 
invariance of the Noether charge becomes trivial only after the corresponding 
mathematical machinery is fully developed \cite{35}. However, all this does 
not make the Lagrangian viewpoint on symmetries obsolete, especially in field 
theory where it proved  to be very useful being ``one of the basic building 
blocks of modern field theories'' \cite{1A}. 

``Judging by the number of papers devoted to it, Noether's theorem must be one 
of the most popular propositions of all time'' \cite{47}. Of course the 
simplicity of the Hamiltonian proof of the Noether theorem, without 
appreciating the rich underlying geometrical structures, cannot explain this
popularity of the Noether theorem and if naively presented can only obscure its
real significance. In fact to call its Hamiltonian counterpart ``the Noether 
theorem'' is not quite correct because ``the Hamiltonian point of view appears 
nowhere in Noether's work, and it is therefore inappropriate to give her name 
to this important, yet easily proved result'' \cite{44}.

\appendix*
\section{Calculation of the commutator $[\hat G+\tau^\nu_{,\,\nu},\,
\hat N^\mu]$}
We have
\begin{equation}
[\hat G+\tau^\nu_{,\,\nu},\,\hat N^\mu]=\hat G(\tau^\mu)+\hat G(\sigma_a)\,
\frac{\partial}{\partial u_{a,\,\mu}}+\sigma_a\left [\hat G,\,\frac{\partial}
{\partial u_{a,\,\mu}}\right]+\sigma_a\left [\tau^\nu_{,\,\nu},\,
\frac{\partial}{\partial u_{a,\,\mu}}\right].
\label{A1}
\end{equation}
But
\begin{equation}
\left [\tau^\nu_{,\,\nu},\,\frac{\partial}{\partial u_{a,\,\mu}}\right]=
-\frac{\partial \tau^\nu_{,\,\nu}}{\partial u_{a,\,\mu}}=-\frac{\partial 
\tau^\mu}{\partial u_a},
\label{A2}
\end{equation}
because
\begin{equation}
\tau^\nu_{,\,\nu}=\frac{\partial \tau^\nu}{\partial x^\nu}+u_{b,\,\nu}\,
\frac{\partial \tau^\nu}{\partial u_b}.
\label{A3}
\end{equation}
On the other hand, as $\tau^\mu$ doesn't depend on field derivatives,
\begin{equation}
\hat G(\tau^\mu)=\tau^\nu\,\tau^\mu_{,\,\nu}+\sigma_a\,\frac{\partial 
\tau^\mu}{\partial u_a}.
\label{A4}
\end{equation}
Further we have
\begin{equation}
\hat G(\sigma_a)=\tau^\nu\,\sigma_{a,\,\nu}+\sigma_b\,\frac{\partial 
\sigma_a}{\partial u_b}+\sigma_{b,\,\nu}\,\frac{\partial \sigma_a}
{\partial u_{b,\,\nu}}.
\label{A5}
\end{equation}
But 
$$\sigma_a=\xi(x,u)-\tau^\mu(x,u)\,u_{a,\,\mu},$$
and, therefore,
\begin{equation}
\frac{\partial \sigma_a}{\partial u_{b,\,\nu}}=-\delta_a^b\,\tau^\nu.
\label{A6}
\end{equation}
Substituting this into (\ref{A5}), we get
\begin{equation}
\hat G(\sigma_a)=\sigma_b\,\frac{\partial \sigma_a}{\partial u_b}.
\label{A7}
\end{equation}
It remains to calculate the commutator
\begin{equation}
\left [\hat G,\,\frac{\partial} {\partial u_{a,\,\mu}}\right]=
\left [\tau^\nu \,\frac{\partial }{\partial x^\nu}+\xi_b\,\frac{\partial }
{\partial u_b}+\eta_{b\nu}\,\frac{\partial }{\partial u_{b,\,\nu}},\,
\frac{\partial} {\partial u_{a,\,\mu}}\right]=-\frac{\partial \eta_{b\nu}}
{\partial u_{a,\,\mu}}\,\frac{\partial }{\partial u_{b,\,\nu}},
\label{A8}
\end{equation}
where we have used the fact that $\tau^\nu$ and $\xi_a$ do not depend on
field derivatives. Using
$$\eta_{b\nu}=\xi_{b,\,\nu}-\tau^\alpha_{,\,\nu}\,u_{b,\,\alpha},$$
along with
\begin{equation}
\frac{\partial \xi_{b,\,\nu}}{\partial u_{a,\,\mu}}=\delta^\mu_\nu\,
\frac{\partial \xi_b}{\partial u_a},\;\;\;\frac{\partial 
\tau^\alpha_{,\,\nu}}{\partial u_{a,\,\mu}}=\delta^\mu_\nu\,
\frac{\partial \tau^\alpha}{\partial u_a},
\label{A9}
\end{equation}
we get
\begin{equation}
\frac{\partial \eta_{b\nu}}{\partial u_{a,\,\mu}}=\delta^\mu_\nu\,\left (
\frac{\partial \xi_b}{\partial u_a}-u_{b,\,\alpha}\,\frac{\partial 
\tau^\alpha}{\partial u_a}\right )-\delta_a^b\,\tau^\mu_{,\,\nu}=
\delta^\mu_\nu\,\frac{\partial \sigma_b}{\partial u_a}-\delta_a^b\,
\tau^\mu_{,\,\nu}.
\label{A10}
\end{equation}
Therefore
\begin{equation}
\left [\hat G,\,\frac{\partial} {\partial u_{a,\,\mu}}\right]=
\tau^\mu_{,\,\nu}\,\frac{\partial }{\partial u_{a,\,\nu}}-
\frac{\partial \sigma_b}{\partial u_a}\,\frac{\partial }
{\partial u_{b,\,\mu}}.
\label{A11}
\end{equation}
Now (\ref{A2}), (\ref{A4}), (\ref{A7}) and (\ref{A11}), in combination with
(\ref{A1}), imply the desired result (\ref{eq65}):
\begin{equation}
[\hat G+\tau^\nu_{,\,\nu},\,\hat N^\mu]=\tau^\mu_{,\,\nu}\,\left (\tau^\nu+
\sigma_a\,\frac{\partial }{\partial u_{a,\,\nu}}\right )=\tau^\mu_{,\,\nu}\,
\hat N^\nu.
\nonumber
\end{equation}

\begin{acknowledgments}
Igor Khavkine  is acknowledged for drawing our attention to Olver's 
Proposition 5.64. The work is supported by the Ministry of Education and 
Science of the Russian Federation and in part by Russian Federation President 
Grant for the support of scientific schools NSh-2479.2014.2. 
\end{acknowledgments}

\end{document}